# Causality between investor sentiment and the shares return on the Moroccan and Tunisian financial markets

# Causalité entre le sentiment des investisseurs et le rendement des actions sur les marchés financiers marocain et tunisien

Dr. CHNIGUIR Mounira

chniguir_m@yahoo.fr

Pr. Jamel Eddine HENCHIRI

Jamel.Henchiri@isggb.rnu.tn

RED-ISGG

University of Gabès, TUNISIA

**Date de soumission** : 05/09/2022

**Date d'acceptation** : 08/12/2022

Pour citer cet article :

CHNIGUIR M. & HENCHIRI J.E. (2022) « Causality between investor sentiment and the shares return on the Moroccan and Tunisian financial markets », Revue du contrôle, de la comptabilité et de l'audit « Volume 6 : numéro 4 » pp : 350 - 364

**Digital Object Identifier :** https://doi.org/10.5281/zenodo.7780959

**Résumé**
Cet article vise à tester la relation entre le sentiment des investisseurs et le rendement des actions cotées sur deux marchés financiers émergents, le marocain et le tunisien. Deux mesures indirectes du sentiment des investisseurs sont utilisées, SENT et ARMS. Ces indicateurs de sentiment montrent qu'il existe une relation importante entre les rendements des actions et le sentiment des investisseurs. Les résultats de la modélisation montrent que le sentiment a une mémoire faible. Par ailleurs, les séries de changements de sentiment ont une mémoire significative. Les résultats du test de causalité de Granger entre le rendement boursier et le sentiment des investisseurs nous montrent que la rentabilité cause le sentiment des investisseurs et non l'inverse pour les deux marchés financiers étudiés.

Grâce à quatre relations autorégressives estimées entre le sentiment des investisseurs, le changement de sentiment, le rendement des actions et le changement de rendement des actions, nous constatons d'un côté que les rendements prédisent les changements de sentiments ce qui confirme notre hypothèse et d'un autre côté que la variation de la rentabilité affecte négativement le sentiment des investisseurs.

Nous concluons que, quelle que soit la mesure du sentiment utilisée, il existe une relation positive et significative entre le sentiment des investisseurs et la rentabilité, mais les rendements ne peuvent pas être prédit à partir de nos diverses variables.

**Mots-clés :** corrélation, prédiction, sentiment des investisseurs, Granger, VAR, Tunindex, MASI

**JEL : G11, G14, N27, O16.**

**Abstract**
This paper aims to test the relationship between investor sentiment and the profitability of stocks listed on two emergent financial markets, the Moroccan and Tunisian ones. Two indirect measures of investor sentiment are used, SENT and ARMS. These sentiment indicators show that there is an important relationship between the stocks returns and investor sentiment. Indeed, the results of modeling investor sentiment by past observations show that sentiment has weak memory; on the other hand, series of changes in sentiment have significant memory. The results of the Granger causality test between stock return and investor sentiment show us that profitability causes investor sentiment and not the other way around for the two financial markets studied.

Thanks to four autoregressive relationships estimated between investor sentiment, change in sentiment, stock return and change in stock return, we find firstly that the returns predict the changes in sentiments which confirms with our hypothesis and secondly, the variation in profitability negatively affects investor sentiment.

We conclude that whatever sentiment measure is used there is a positive and significant relationship between investor sentiment and profitability, but sentiment cannot be predicted from our various variables.

**Keywords :** correlation, prediction, investor sentiment, Granger, VAR, Tunindex, MASI
**JEL: G11, G14, N27, O16.**

## Introduction

Several empirical studies have shown that investor sentiment is an explanatory variable for equity returns. In this context, Kumar and lee (2006) found that individual investor sentiment has significant explanatory power for returns.

After having demonstrated in previous work that there is an important relationship between investor sentiment and stock return for the two Maghreb markets (Morocco and Tunisia), we ask ourselves here, as Neal et al. (1998) then Kumari & Mahakud (2015), if sentiments predict the stock return of our markets and which still remains unanswered.

### 1. investor sentiment and equity returns on the financial markets

The notion of investor sentiment was introduced by Nicholas Barberis, Andrei Shleifer and Robert Vishny (1998). Fisher & Statman (2000) then Baker & Wurgler (2007) studied the relationship between investor sentiment and stock performance in financial markets.

Brown and Cliff (2004) show that there is a negative Sentiment-Profitability relationship on the level of the US stock market and find a difference between the Profitability / Sentiment relationship and the Profitability / Sentiment level relationship while Kahneman and Tversky (1979) prove that the investor is sensitive to the change in wealth and not to total wealth.

Brown and Cliff (2005) found no correlation between investor sentiment and future short-term stock performance and a negative correlation between the two variables for the medium term.

To verify the robustness of the relationship between investor sentiment and return and to study the direction of information flow between these variables, Granger and Auto Regressive Vector VAR causality methodologies were used.

Our hypotheses are to state that there is a relationship between investor sentiment and the stock return of stocks and so, this sentiment has a predictive power on future returns.

Granger's causality test (1969) examines whether variation in returns is the cause of change in investor sentiment, and vice versa. This test is designed to determine whether the two-time series pass one after the other or simultaneously. This test verifies the dynamic relationship between yield and volume. The optimal number of delays was determined based on the Akaïke information criterion.

Thus, sentiment indicators are excellent signals for trading on the stock market. Indeed, behavioral finance claims that investors tend to overvalue prices when they are pessimistic.

As a result, as these temporary deviations will be corrected and as securities prices converge towards their fundamentals, current investor optimism will be negatively correlated with future returns. So, the more optimistic (pessimistic) investors are about the future performance of the markets, the less (more) future returns will be good.

Besides the theoretical framework, empirical analysis of the dot-com bubble in the late 1990s suggests that returns and sentiment indicators can act as a system. When the market turned bullish, sentiment indicators hit record highs (there could be a feedback mechanism between sentiment and returns).

In short, it is a question of whether investor sentiment predicts equity returns or rather the other way around.

In addition, we estimate a VAR model to examine the causality between sentiment and returns of stocks, i.e., this model allows us to examine on the one hand how returns and sentiment interact and the sense of causality using Granger's test on the other hand.

The VAR model is presented as follows:

$$Y_t = \mu + \sum_{i=0}^{P} \emptyset i \, Y_{t-i} + \varepsilon_t$$

as:

$Y_t$, $Y_{t-i}$: the vectors of stock return and investor sentiment on the date t & t-i ;

$\mu$: the vector of constant terms;

$\emptyset$: the matrix of autoregressive coefficients;

$\varepsilon_t$: the vector of the residuals of the regression;

p: the optimal number of shifts determined by minimizing the Akaïke information criterion (P = 2 in our case).

## 2. Data, measurements and methodology

Our sample used to perform this analysis consists of the daily closing prices of shares listed on the Casablanca Stock Exchange (Casa SE or BVC) and the Tunis Stock Exchange (TuSE or BVMT) for the period from January 1, 2006, to 31 December 2015.

The choice of these two countries amounts to putting in opposition a country which experienced a revolution in 2011 which had a strong impact on its economic activity and a nation which chose to adopt gentle changes and therefore to preserve its socio-economic foundations. In addition, there are many points in common between these two countries which are the sharing of their affiliation (North Africans), their historical and cultural status, and their strong dependence on Europe.

In fact, for this analysis we need stock prices, stock indexes, Tunisian three-month treasury bill rate, market capitalizations of listed companies that were collected from the Tunis and Casablanca stock exchange website.

The returns $r_t$ are calculated by the usual arithmetic formula. Descriptive statistics are available from the authors.

The variable sent is calculated as follows:

$$sent_1 = \frac{ADV_t}{DEC_t}$$

And the variable ARMS is calculated as follows:

$$ARMSt = \frac{ADVt/ADV\ volt}{DECt/DEC\ vol\ t}$$

### 3. Causality between investor sentiment and return on assets

Our study is based on a VAR methodology, extended by a search for causal links. The identification of causal relationships between stock return and volume allows a better apprehension and understanding of psychological and financial phenomena and provides additional information about the anteriority of events between them. The analysis of causality à la Granger can therefore be carried out from the estimation of a VAR model and the application of the causality test following the example of Chuang and Lee (2006) and Statman, Thorley and Vorkink (2006). The analysis can be continued on a classical modeling which said without error correction since the series are stationary.

In fact, the Granger function can be written in the following form:

$$\begin{bmatrix} rt_t \\ sent_t \end{bmatrix} = \begin{bmatrix} \alpha_{rt,t} \\ \alpha_{sent,t} \end{bmatrix} + \sum_{i=1}^{r} \begin{bmatrix} \beta_{rt,t} \\ \beta_{sent,t} \end{bmatrix} \begin{bmatrix} rt_{t-i} \\ sent_{t-i} \end{bmatrix} + \varepsilon_t$$

The results of the Granger causality test are given by the following Tables 1 & 2:

| Table 1: causality test between stock return and sentiment: Tunis SE | | | | |
|---|---|---|---|---|
| | Lag1 | | Lag2 | |
| | Test1 | Test2 | Test1 | Test2 |
| SENT | **1E-159** | 0.8328 | **4E-158** | 0.9729 |
| ARMS | **5E-11** | 0.1153 | **6E-10** | 0.3990 |
| Test 1: stock return causes sentiment | | | | |
| Test2: sentiment causes stock return | | | | |
| P-value<0.05: meaning of causality is significant | | | | |

| Table 2: causality test between stock return and sentiment: Casa SE | | | | |
|---|---|---|---|---|
| | Lag1 | | Lag2 | |
| | Test1 | Test2 | Test1 | Test2 |
| SENT | 0.9516 | 0.3145 | **0.034** | 0.5158 |
| ARMS | **0.0067** | 0.5781 | **000268** | 0.340 |
| Test 1: stock return causes sentiment | | | | |
| Test 2: sentiment causes stock return | | | | |
| P-value<0.05: meaning of causality is significant | | | | |

The results of the Granger causality test between stock return and investor sentiment show us that stock return causes investor sentiment and not the other way around for the two financial markets studied.

In fact, we find 4 tests performed that validate the sense that stock return causes sentiment: two tests with 2 delays and two tests with a single delay.

This validation is due to a value of p-value <0.05.

On the other hand, for test 2, which seek if the sentiment causes stock return, no test validates this sense of causality.

In the same way for the Moroccan market, one notes that 3 tests carried out validate the direction that stock return causes the sentiment: two tests with 2 delays and one test with a delay. This is equivalent to the results found for the Tunisian market.

In addition, to validate our findings, we estimate the coefficients of two-way autoregressive vectors. The results of this estimate are given in tables 3 & 4 below:

| Table 3: VAR (Stock return & sentiment measures) : Tunis SE | | | | | |
|---|---|---|---|---|---|
| | RT | SENT | | RT | ARMS |
| RT (-1) | 0.0369 | -1.9222 | RT (-1) | 0.0282 | 0.2707 |
| | **(1.8825)** | (-0.2246) | | (1.4414) | (0.7979) |
| RT (-2) | 0.0371 | -0.4424 | RT (-2) | 0.0423 | 2.9477 |
| | **(2.1752)** | (-0.059) | | **(2.1565)** | (0.019) |
| SENT (-1) | 0.0013 | -0.0007 | ARMS (-1) | 0.0353 | 0.0349 |
| | (28.908) | (-0.036) | | **(1.80116)** | **(1.7825)** |
| SENT (-2) | -4.99E-05 | 0.0019 | ARMS (-2) | 0.0298 | 0.0050 |
| | (-0.9606) | (0.0864) | | (0.920) | (0.2546) |
| C | -0.0496 | 2.0910 | C | -0.0475 | 0.0542 |
| | (-15.553) | (1.5036) | | (-13.199) | (0.8720) |

* Significant at 5% level

| Table 4: VAR (Stock return & sentiment measures): Casa SE | | | | | |
|---|---|---|---|---|---|
| | RT | SENT | | RT | ARMS |
| RT (-1) | 0.0381 | -1.9222 | RT (-1) | 0.0381 | 0.2339 |
| | **(1.9482)** | (-0.2246) | | (1.234) | (0.1372) |
| RT (-2) | 0.0371 | -0.4424 | RT (-2) | 0.001 | 0.2604 |
| | **(2.1752)** | (-0.059) | | **(2.6519)** | (0.0136) |
| SENT (-1) | 0.0375 | -0.0007 | ARMS (-1) | 0.0637 | 0.1601 |
| | (0.2665) | (-0.036) | | **(1.8372)** | **(8.1947)** |
| SENT (-2) | -0.00029 | 0.03816 | ARMS (-2) | 0.01014 | 0.0050 |
| | (-10770) | (0.0195 | | (0.5176) | (0.2546) |
| C | 0.0052 | 1.464 | C | 0.864 | 3.25 |
| | (1.1440) | (10.336) | | (15.634) | (1.1263) |

*Significant at 5% level

According to the tables above, whether for the Tunisian or Moroccan market, and for the vectors of returns, the coefficients of the delayed returns are significant, hence the daily returns have a memory. So that it is possible to predict present returns from past returns. However, the coefficients of the delayed sentiments are insignificant so we cannot predict the present returns from the past sentiments.

For the sentiment vector, the lagged coefficients are insignificant, which lead us to conclude that we cannot predict the present sentiments either from past returns or from past sentiments.

### 4. Causality between stock return and change in investor sentiment

The aim is to study the relationship between the variation of two indicators of sentiment and the stock return.

$$\begin{bmatrix} rt_t \\ \Delta sent_t \end{bmatrix} = \begin{bmatrix} \alpha_{rt,t} \\ \alpha_{sent,t} \end{bmatrix} + \sum_{i=1}^{r} \begin{bmatrix} \beta_{rt,t} \\ \beta_{sent,t} \end{bmatrix} \begin{bmatrix} rt_{t-i} \\ \Delta sent_{t-i} \end{bmatrix} + \varepsilon_t$$

The Granger test causality result is provided by the following Tables 5 & 6:

| Table 5: causality test between stock return & sentiment change: Tunis SE | | | | |
|---|---|---|---|---|
| | Lag1 | | Lag2 | |
| | Test1 | Test2 | Test1 | Test2 |
| ΔSENT | **0.0023** | 0.098 | **0.0102** | 0. 8959 |
| ΔARMS | **2E-05** | 0.3071 | **2E-07** | 0. 4369 |
| Test 1: stock return causes sentiment change | | | | |
| Test 2: sentiment change causes stock return | | | | |
| P-value<0.05: meaning of causality is significant | | | | |

| Table 6: causality test between stock return & sentiment change: Casa SE | | | | |
|---|---|---|---|---|
| | Lag1 | | Lag2 | |
| | Test1 | Test2 | Test1 | Test2 |
| SENT | 0.8439 | 0.9806 | **4E-158** | 0.9729 |
| ARMS | **5E-11** | 0.1153 | **6E-10** | 0.3990 |
| Test 1: stock return causes sentiment change | | | | |
| Test 2: sentiment change causes stock return | | | | |
| P-value<0.05: meaning of causality is significant | | | | |

The results of the causal test between return and change in investor sentiment show that there is a strong relationship, and that profitability causes the change in sentiment.

Thus, we find on 4 tests carried out, 3 tests which approve the sense that stock return causes the sentiment variation: two tests with two delays and one test with a single delay. This validation is due to a P-value <0.05.

For the second sense of causation, i.e., the change in investor sentiment causes stock return, no test validates this sense of causation.

Finally, these findings are similar for the Moroccan market.

In fact, and to validate our findings (stock return causes the variation in sentiment), we estimate the coefficients of the two-way autoregressive vectors. The results of this estimate are given by this table 7 for the Tunisian market and table 8 for the Moroccan market:

| Table 7: VAR (Stock return & sentiment measures) : Tunis SE | | | | | |
|---|---|---|---|---|---|
| | RT | SENT | | RT | ARMS |
| RT (-1) | 0.028 | 0.0023 | RT (-1) | 0.0573 | -2.882 |
| | (1.4629) | (0.025) | | (1.8925) | (-1.0741) |
| RT (-2) | 0.035 | 0.0447 | RT (-2) | 0.0935 | -2.94965 |
| | **(1.8065)** | (0.4673) | | **(3.0839)** | (-1.824) |
| ΔSENT (-1) | 0.0013 | 0.040 | ΔARMS (-1) | 0.0014 | -0.5126 |
| | (28.908) | **(2.0373)** | | **(4.2723)** | **(16.8436)** |
| ΔSENT (-2) | -4.99E-05 | 0.2019 | ΔARMS (-2) | 0.0012 | -0.3132 |
| | (-0.9606) | (0.0824) | | (3.6925) | (-10.214) |
| C | 0.001 | -0.064 | C | 0.0001 | 0.0023 |
| | (0.142) | (-0.0466) | | (0.072) | (0.8760) |

| Table 8: VAR (Stock return & sentiment measures): Casa SE | | | | | |
|---|---|---|---|---|---|
| | RT | SENT | | RT | ARMS |
| RT (-1) | 0.010177 | -1.9222 | RT (-1) | 0.0282 | -0.345567 |
| | (0.51917) | (-0.2246) | | (1.4414) | (-18.7704) |
| RT (-2) | 0.0795 | -0.4424 | RT (-2) | 0.0064 | -7.11E-05 |
| | **(2.6155)** | (-0.059) | | (0.3913) | (-0.26535) |
| ΔSENT (-1) | 0.030051 | -0.0007 | ΔARMS (-1) | 0.0372 | 0.2678 |
| | (0.18383) | (-0.036) | | **(2.2569)** | (1.7426) |
| ΔSENT (-2) | 0.003354 | 0.0019 | ΔARMS (-2) | 0.1899 | 0.0050 |
| | (0.17109) | (0.0864) | | (1.5235) | (0.2546) |
| C | 0.000996 | -0.288295 | C | -0.0372 | 0.6931 |
| | (0.44264) | (-15.3605) | | (-1.7635) | (5.2148) |

Note that for the stock return vectors, the delayed stock return coefficients are significant, which means that the daily returns have a memory. Likewise, the coefficients of the delays of the variations in sentiment are significant, which proves that the change in daily sentiment has a memory of two days.

For the vectors of the sentiment change, we see that the coefficients are significant; therefore, the sentiment change affect the returns.

## 5. Causality between the variation in stock return and investor sentiment

Kahneman and Tversky (1979) prove that the investor is sensitive to the change in wealth and not to total wealth, in other words investor sentiment is not sensitive to total wealth but to the change in wealth. In this context, it is interesting to test the relationship between the variation in profitability and investor sentiment.

The Granger causality test is written in the following form:

$$\begin{bmatrix} \Delta rt_t \\ sent_t \end{bmatrix} = \begin{bmatrix} \alpha_{rt,t} \\ \alpha_{sent,t} \end{bmatrix} + \sum_{i=1}^{r} \begin{bmatrix} \beta_{rt,t} \\ \beta_{sent,t} \end{bmatrix} \begin{bmatrix} \Delta rt_{t-i} \\ sent_{t-i} \end{bmatrix} + \varepsilon_t$$

The results of this test are provided by Tables 9 and 10 below:

| Table 9: stock return change & sentiment measures: Tunis SE | | | | |
|---|---|---|---|---|
| | Lag1 | | Lag2 | |
| | Test1 | Test2 | Test1 | Test2 |
| SENT | 0.9176 | **2E-106** | **2E-136** | 0.9910 |
| ARMS | 0.7611 | **0.0461** | 0.8889 | **0. 0152** |
| Test 1: stock return change causes sentiment. | | | | |
| Test 2: sentiment causes stock return change | | | | |
| P-value<0.05: meaning of causality is significant | | | | |
| Table 10: stock return change & sentiment measures: Casa SE | | | | |
| | Lag1 | | Lag2 | |
| | Test1 | Test2 | Test1 | Test2 |
| SENT | **2E-05** | 0.3071 | **0.0102** | 0. 8959 |
| ARMS | **2E-05** | 0.3071 | **2E-07** | 0.6943 |
| Test 1: stock return change causes sentiment. | | | | |
| Test 2: sentiment causes stock return change | | | | |
| P-value<0.05: meaning of causality is significant | | | | |

About the first direction of causality, the stock return change causes the sentiment measures, we find on 4 tests carried out, a single test which validates this direction of causality, that there is no direction of causality between variation in stock return and sentiment measures.

For the second sense of causality, i.e., measures of sentiment affect change in stock return, 3 tests were found that validate that sentiment causes change in stock return.

To confirm our findings, we estimate the coefficients of the two-way autoregressive vectors. The results of this estimate are given in tables 11 and 12 below:

| Table 11: VAR (Stock return & sentiment measures): Tunis SE | | | | | |
|---|---|---|---|---|---|
| | RT | SENT | | RT | ARMS |
| ΔRT (-1) | -0.439 | 3.784 | ΔRT (-1) | -0.446 | -0.345567 |
| | **(-14.548)** | (0.473) | | (-14.21) | (-18.7704) |
| ΔRT (-2) | -0.134 | 0.982 | ΔRT (-2) | -0.128 | 1.177 |
| | **(-4.544)** | (0.125) | | **(-4.120)** | (0.591) |
| SENT (-1) | 2.42E-05 | -0.015 | ARMS (-1) | -0.001 | 0.0372 |
| | (0.2) | (-0.475) | | **(-2.057)** | (2.2569) |
| SENT (-2) | -0.001 | -0.006 | ARMS (-2) | -0.0001 | 0.043 |
| | (-8.873) | (-0.213) | | (-0.237) | (1.357) |
| C | 0.0001 | 1.464 | C | 0.001 | 0.001 |
| | (2.125) | (10.336) | | (1.215) | (1.415) |

| Table 12: VAR (Stock return & sentiment measures): Casa SE | | | | | |
|---|---|---|---|---|---|
| | RT | SENT | | RT | ARMS |
| RT (-1) | 0.4895 | 0.8635 | RT (-1) | -0.2392 | 0.2246 |
| | **(-4.0914)** | (0.8680) | | (-1.866) | (1.3797) |
| RT (-2) | -0.2807 | -0.3373 | RT (-2) | -0.7005 | 0.1977 |
| | **(-2.2761)** | (-0.3403) | | **(-4.3212)** | (1.2885) |
| ΔSENT (-1) | 0.0119 | 0.1599 | ΔARMS (-1) | -0.4111 | 0.5685 |
| | (0.0.9333) | (1.3030) | | **(2.8647)** | (3.4065) |
| ΔSENT (-2) | -0.0174 | 0.1750 | ΔARMS (-2) | 0.0043 | -0.8218 |
| | (-0.9219) | (0.1511) | | (0.2497) | (-0.7722) |
| C | 0.002 | 0.6133 | C | 0.0192 | 0.6260 |
| | (0.0216) | (4.8276) | | (0.9481) | (3.8443) |

For the vectors of the stock return changes, the lagged coefficients of the stock return variations are significant; this implies that the stock return change has a memory. Thus, the sentiment lag coefficients are overall negatively significant and therefore the changes in stock return negatively influences investor sentiment.

For sentiment vectors, the lagged coefficients of changes in returns and sentiment measures are insignificant, allowing us to conclude that sentiment cannot be predicted from changes in past returns and sentiment.

## 6. Causality between the variation in stock return and the variation in sentiment

We will study here the relationship between change in sentiment and change in stock return. So, we perform the Granger causality test of the following form:

$$\begin{bmatrix} \Delta rt_t \\ \Delta sent_t \end{bmatrix} = \begin{bmatrix} \alpha_{rt,t} \\ \alpha_{sent,t} \end{bmatrix} + \sum_{i=1}^{r} \begin{bmatrix} \beta_{rt,t} \\ \beta_{sent,t} \end{bmatrix} \begin{bmatrix} \Delta rt_{t-i} \\ \Delta sent_{t-i} \end{bmatrix} + \varepsilon_t$$

The results of this test are summarized in the following table:

| | Table 13: stock return change & sentiment change: Tunis SE | | | |
|---|---|---|---|---|
| | Lag1 | | Lag2 | |
| | Test1 | Test2 | Test1 | Test2 |
| SENT | 0.5378 | 0.6336 | 0.9220 | **0.0109** |
| ARMS | 0.4049 | **0.0091** | 0.9857 | 0.7167 |
| Test 1: stock return change causes sentiment change | | | | |
| Test 2: sentiment change causes stock return change | | | | |
| P-value<0.05: meaning of causality is significant | | | | |

| | Table 14: stock return change & sentiment change: Casa SE | | | |
|---|---|---|---|---|
| | Lag1 | | Lag2 | |
| | Test1 | Test2 | Test1 | Test2 |
| SENT | 0.8563 | 0.7413 | 0.1527 | **0.0475** |
| ARMS | 0.9247 | 0.9477 | 0.9836 | 0. 9191 |
| Test 1: stock return change causes sentiment change | | | | |
| Test 2: sentiment change causes stock return change | | | | |
| P-value<0.05: meaning of causality is significant | | | | |

For the first sense of causality, there is no test that validates this sense; hence the change in stock return does not causes the change in sentiment.

For the second sense of causality, i.e., the variation in sentiment causes the variation in stock return, two tests which validate this direction, so the sentiment change causes stock return change.

To validate our results, we estimate the coefficients of the autoregressive vectors in both directions. The results of this estimate are provided by the table below:

| Table 15: VAR (stock return change & sentiment change): Tunis SE ||||||
|---|---|---|---|---|---|
| | ΔRT | ΔSENT | | ΔRT | ΔARMS |
| ΔRT (-1) | -0.441 | -3.256 | ΔRT (-1) | -0.444 | -1.59 |
| | **(-14.437)** | (-0.352) | | **(-14.02)** | (-0.829) |
| ΔRT (-2) | -0.131 | -0.096 | ΔRT (-2) | -0.127 | 0.069 |
| | **(-4.471)** | (-0.01) | | **(-4.1)** | (0.036) |
| ΔSENT (-1) | 0.0003 | -0.096 | ΔARMS (-1) | 0.0002 | -0.502 |
| | **(3.614)** | **(-22.565)** | | 0.649) | **(-16.404)** |
| ΔSENT (-2) | 0.0007 | -0.0343 | ΔARMS (-2) | -0.0003 | -0.306 |
| | **(7.255)** | **(-11.482)** | | (-0.8) | **(-9.978)** |
| C | -0.0001 | -0.006 | C | 0.003 | 0.002 |
| | (-0.745) | (-0.044) | | (0.763) | (0.075) |

| Table 16: VAR (stock return change & sentiment change) : Casa SE ||||||
|---|---|---|---|---|---|
| | ΔRT | ΔSENT | | ΔRT | ΔARMS |
| ΔRT (-1) | -0.7128 | 0.8706 | ΔRT (-1) | 0.010 | -1.4279 |
| | **(-6.059)** | (0.8057) | | (0.4574) | (-1.1212) |
| ΔRT (-2) | -0.2641 | -0.1424 | ΔRT (-2) | -0.7818 | -0.9642 |
| | **(-2.2526)** | (-0.1322) | | **(-4.632)** | (-0.7499) |
| ΔSENT (-1) | -0.0104 | -0.8365 | ARMS (-1) | 0.0128 | -0.3705 |
| | (-0.8349) | (-7.3225) | | (0.5586) | **(-2.1512)** |
| ΔSENT (-2) | -0.0191 | -0.3863 | ARMS (-2) | 0.0238 | -0.5459 |
| | (-1.5397 | (-3.3771)) | | (1.4734) | **(-3.2718)** |
| C | 0.0008 | -0.001 | C | 0.0007 | 0.0081 |
| | (0.1931) | (0.0256) | | (0.166) | (0.2351) |

For the vectors of the variations in returns, the lagged coefficients of returns change are significant, which indicates that the variation in daily returns has a memory. The coefficients of changes in delayed sentiment are insignificant, so we cannot predict changes in returns from changes in past sentiment.

For the vectors of variations in sentiment, the lagged coefficients are significant, which implies that the variation in sentiment causes the variation in profitability.

**Conclusion**

We tested the relationship between investor sentiment and the profitability of stocks listed on the Tunisian and Moroccan financial markets. Two indirect measures of investor sentiment are used, the first measure is the SENT indicator which is the ratio of the number of securities that have experienced an increase divided by those that have experienced a decrease, and the second indicator is the measure ARMS computed by the number of securities that experienced a price *increase* standardized by their trading volumes divided by the number of securities that experienced a price *decrease* standardized by the trading volumes of those securities. These sentiment indicators show that there is an important relationship between the profitability of stocks and investor sentiment.

Indeed, the results of modeling investor sentiment by past observations show that sentiment has weak memory; on the other hand, series of changes in sentiment have significant memory. Next, four estimated autoregressive relationships between investor sentiment, sentiment change, profitability, and profitability change. First, we found that the returns predict changes in sentiment, which confirms with our hypothesis, and second, the change in profitability negatively affects investor sentiment.

We conclude that whatever sentiment measure is used there is a positive and significant relationship between investor sentiment and equity returns, this result is similar to the results of Brown and Cliff (2004) but we cannot predict returns from our variables.